%% file: main.tex
\documentclass[journal]{IEEEtran}
\usepackage{cite}
\usepackage[pdftex]{graphicx}
\usepackage{svg}
\usepackage{algorithmic}
\usepackage{array}
\usepackage[caption=false,font=normalsize,labelfont=sf,textfont=sf]{subfig}
\usepackage{stfloats}

\input{standard.tex}
\makeatletter
\def\section{\@startsection{section}{1}{\z@}{-3.0ex plus -1.5ex minus -1.5ex}
{0.7ex plus 1ex minus 0ex}{\normalfont\normalsize\centering\scshape}}%
\def\subsection{\@startsection{subsection}{2}{\z@}{-3.5ex plus -1.5ex minus -1.5ex}%
{0.7ex plus .5ex minus 0ex}{\normalfont\normalsize\itshape}}%
\makeatother

\newenvironment{sproof}{%
  \proof}{\endproof}

\begin{document}

\title{Sequential One-Sided Hypothesis Testing of Markov Chains}

\author{Greg~Fields,~
        Tara~Javidi,~
        and~Shubhanshu~Shekhar

\thanks{G. Fields and T. Javidi are with ECE UC San Diego.}
\thanks{S. Shekhar is with EECS University of Michigan, Ann Arbor.}
\thanks{Manuscript received June 30, 2024}}

\markboth{}%
{Shell \MakeLowercase{\textit{et al.}}: Sequential One-Sided Hypothesis Testing of Markov Chains}

\maketitle

\begin{abstract}
    We study the problem of sequentially testing whether a given stochastic process is generated by a known Markov chain. 
    Formally, given access to a stream of random variables, we want to quickly determine whether this sequence is a trajectory of a Markov chain with a known transition matrix $P$ (null hypothesis) or not (composite alternative hypothesis). 
    This problem naturally arises in many engineering problems. 

    The main technical challenge is to develop a sequential testing scheme that adapts its sample size to the unknown alternative. Indeed, if we knew the alternative distribution (that is, the transition matrix) $Q$, a natural approach would be to use a generalization of Wald's sequential probability ratio test (SPRT).  Building on this intuition, we propose and analyze a family of one-sided SPRT-type tests for our problem that use a data-driven estimator $\hat{Q}$. In particular, we show that if the deployed estimator admits a worst-case regret guarantee scaling as $\mc{O}\lp \log{t} \rp$, then the performance of our test asymptotically matches that of SPRT in the simple hypothesis testing case. In other words, our test automatically adapts to the unknown hardness of the problem, without any prior information. We end with a discussion of known Markov chain estimators with $\mc{O}\lp \log{t} \rp$ regret.     
\end{abstract}


\section{Introduction}
\label{sec:intro}

\input{10_introduction}


\section{Problem Setup}
In this section, we provide the mathematical formulation of our problem and provide an overview of our contributions over related work. 
\label{sec:problem}
\input{20_problem-formulation}


\section{Theoretical Results}
\label{sec:theory}
\input{30_theory}

\section{Simulations}
\label{sec:simulations}
\input{40_simulations}

\section{Future Work}
\label{sec:future}
\input{50_conclusion}

\section*{Acknowledgment}
We thank the reviewers for their thorough reading of the paper and many useful suggestions and references. Greg Fields and Tara Javidi were supported in part in this project by the  Office of Naval Research (ONR) Award N00014-22-1-2363, industry partners as specified in the Resilient and Intelligent NextG Systems (RINGS) program, and industry partners of the UCSD center for Machine Intelligence, Computing, and Security (MICS). 

\appendices
\input{90_appendix}

\ifCLASSOPTIONcaptionsoff
  \newpage
\fi

\bibliographystyle{IEEEtran}
\bibliography{99_ref}

\end{document}

%% file: standard.tex
\usepackage{xcolor} 
\usepackage{wrapfig}

\usepackage{soul}
\setstcolor{red}

\usepackage{amsmath}
\usepackage{amsthm}
\usepackage{amssymb}
\usepackage{bm} 
\usepackage{bbm} 
\usepackage{mathtools}
\usepackage{enumitem}
\usepackage{nicefrac}
\usepackage{longtable}
\usepackage[boxruled, vlined,linesnumbered]{algorithm2e}
\usepackage{hyperref}

\usepackage{diagbox}
\usepackage{makecell}
\makeatletter
\g@addto@macro\normalsize{%
\setlength\abovedisplayskip{6pt} 
\setlength\belowdisplayskip{6pt}
\setlength\abovedisplayshortskip{6pt} 
\setlength\belowdisplayshortskip{6pt}
}
\makeatother

\newtheorem{theorem}{Theorem}
\newtheorem{lemma}{Lemma}

\newtheorem{remark}{Remark}

\theoremstyle{definition}
\newtheorem{definition}{Definition}

\newcommand{\lp}{\left (} 
\newcommand{\rp}{\right )} 
\newcommand{\lb}{\left [}
\newcommand{\rb}{\right ]}
\newcommand{\lcb}{\left \{}
\newcommand{\rcb}{\right \}} 

\newcommand{\mc}[1]{\mathcal{#1}}

\newcommand{\indi}[1]{\mathbbm{1}_{ \{#1\} }}

\renewcommand{\P}{\mathbb{P}}
\newcommand{\E}{\mathbb{E}}

\DeclarePairedDelimiterX{\norm}[1]{\lVert}{\rVert}{#1}
\DeclarePairedDelimiterX{\inner}[2]{\langle}{\rangle}{#1, #2}

\makeatletter
\newcommand{\pushright}[1]{\ifmeasuring@#1\else\omit\hfill$\displaystyle#1$\fi\ignorespaces}
\newcommand{\pushleft}[1]{\ifmeasuring@#1\else\omit$\displaystyle#1$\hfill\fi\ignorespaces}
\makeatother


%% file: 10_introduction.tex
The problem of monitoring of a stochastic process representing the dynamic state of an environment for the purpose of confirming or rejecting a hypothesis is an age-old problem fundamental to scientific discovery and engineering design. As part of scientific discovery, we often form hypotheses regarding the physical world or biological processes and then collect observations/data to test our theory. Similarly, in many complex engineering problems and in the process of anomaly detection, we monitor an engineered system in order to confirm desirable operation and dynamic behavior. The problem is particularly challenging where the state of the environment is both dynamic and stochastic. For instance, this problem arises naturally in the context of wireless networking: the California Public Utilities Commission’s (CPUC’s) CalSPEED program measures fundamental and dynamic metrics associated with mobile wireless services at about 4,000 locations in urban, rural, and Tribal areas across California. The regulatory monitoring and anomaly detection is technically challenging since the network behavior is both dynamic and subject to inevitable randomness due to the environment. Networks are often designed with some uncertainty tolerated and exhibited in the parameters of the protocol stack such as radio signals' strength and network connections' quality; all subject to stochastic fluctuations. In this example, and other similar problems, a sequence of observations are gathered and tested, providing statistical evidence. 

In this paper, we consider the problem of \textit{sequential composite hypothesis} testing where a tester is given a sequence of samples $X_1,X_2,\ldots, X_t, \ldots$ one-by-one and tasked with deciding if the samples are drawn from a given stochastic process--the null hypothesis--or if they are drawn from a different, unknown process.
At each time step, the tester must decide if more samples are needed or, given the observed samples so far, sufficient evidence has been accumulated to reject the null hypothesis.  Prior work considers the simple case where the observations are independent and identically distributed, whereas we consider a Markov sequence of observations under the null hypothesis. This is motivated by the problem of network resilience and anomaly detection, where we need to explicitly account for temporally correlated dynamics. Furthermore, we consider a \emph{one-sided} testing scenario under which the metric of interest is how fast, under the alternative, the tester can reject the null, while allowing the tester to keep accumulating evidence in favor of the null hypothesis indefinitely. Note that we allow the alternative to be composite: under the alternative, we assume no prior information about the true distribution of the underlying process. 

This sequential one-sided hypothesis testing problem is particularly well-suited to the problem of anomaly detection: when monitoring a process that is functioning within specifications, there is no need to halt the process and confirm it is working as intended.  So, under the null hypothesis, our desired behavior is for the test to continue running indefinitely.  In contrast, if something goes wrong, it is crucial to identify the anomaly and to do so as quickly as possible.  Furthermore, this must happen regardless of the nature of the anomaly; i.e. it is not reasonable to assume the tester has much information about the alternative.  This strongly motivates the \textit{composite} nature of our test, where we test against any deviation from the null hypothesis, instead of requiring a specified alternative hypothesis. Moreover, in anomaly detection we are concerned with rare events across time--a process is functioning properly until it eventually deviates from the norm--this is much more naturally modeled by a stochastic process with memory/state.  

\subsection{Notation}
In general, we will be working on finite state spaces with $m$ states. In the i.i.d. setting we will consider distributions $p\in\Delta_{m-1}$ drawn from the $m-1$ dimensional simplex.  We will identify Markov chains with their transition matrices, $\boldsymbol{P}\in\mc{M}_1^m$, where $\mc{M}_k^m$ denotes the space of $k$th-order ergodic Markov chains on $m$ symbols and $\boldsymbol{P}(j|i)=P_{ij}$ gives the transition probability from state $i$ to state $j$.  

Recall that the usual KL divergence between two discrete distributions $q$ and $p$ is defined as
\begin{align*}
    D(q\parallel p) = \sum_{i=1}^m q_i \log{\frac{q_i}{p_i}}.
\end{align*}
If $\boldsymbol{Q}$ is a Markov chain with stationary distribution $\rho$, then we define a stationary-averaged KL divergence by
\begin{align*}
    D_M(\boldsymbol{Q} \parallel \boldsymbol{P}) \coloneqq \sum_{i=1}^m \rho_i D\big( \boldsymbol{Q}(\cdot|i) {\parallel} \boldsymbol{P}(\cdot|i) \big).
\end{align*}
This quantity weights the KL divergences between the state-transition probabilities of the two chains by how frequently the state will occur under $\boldsymbol{Q}$.  This quantity arises, for example, in error exponent analysis of simple hypothesis testing between two known Markov chains~\cite{natarajan1985largedev}. 

We will use $\P_{H_0}$ and $\E_{H_0}$, to indicate the probability or expectation taken when the null hypothesis is true, i.e. data is generated according to the null hypothesis' distribution.  Finally, we will make use of symmetric Dirichlet distributions--Dirichlet distributions where all parameters are equal--denoted by $D_{(\gamma)}$ for the shared parameter $\gamma>0$. 

%% file: 20_problem-formulation.tex
\subsection{Problem Formulation}
We consider the problem of composite hypothesis testing: a tester is given a known Markov chain, represented by its transition matrix $\boldsymbol{P}\in\mc{M}^m_1$, and a trajectory  $X_1,X_2,\ldots$ drawn from an underlying Markov chain with dynamics $\boldsymbol{P_X}\in\mc{M}^m_1$. 
The tester must decide 
\begin{align*}
    H_0: \boldsymbol{P_X} = \boldsymbol{P}\text{\ \ \ vs \ \ \ \ } H_1: \boldsymbol{P_X} \neq \boldsymbol{P}. 
\end{align*}

 Note that this problem generalizes the classical binary hypothesis testing with i.i.d.\ observations: the composite nature of the alternative hypothesis generalizes \textit{simple} hypothesis testing, where the tester compares against another known distribution. Furthermore, the Markov observation assumption extends the i.i.d.\ case, where the transition matrices reduce to 
 degenerate matrices with all rows equal to $p$ for the known target distribution, and $p_X$ governing the i.i.d.\ samples.  

We work in the sequential setting, where at each time step $t=1,2,\ldots$ the tester may choose to either end the testing and declare a hypothesis or may request another sample, continuing the test.  In this setting, we aim to develop a test which correctly declares $H_1$, rejecting the null hypothesis $\boldsymbol{P}$, in as few samples as possible while controlling type-1 errors.  

\begin{definition}
A \textit{level-$\alpha$, power-one} test is a test which defines a stopping time, $\tau$ with the following guarantees:
\begin{align}
    \label{eq:promises}
    \P_{H_0}(\tau<\infty) \leq \alpha \text{ and } \P_{H_1}(\tau<\infty) = 1. 
\end{align}
\end{definition}
We will characterize the performance of our algorithms via bounds on $\E_{H_1}[\tau]$,  the expected number of samples required to correctly reject the null hypothesis.

\subsection{Related Work}
\label{subsec:related}
Related, but distinct, veriants of hypothesis testing problem have been studied in a wide variety of contexts.
The most important factors that set our work apart from prior work is our general one-sided Markov testing formulation where 1) we consider samples that are not drawn iid in time but instead have a Markovian dynamic, 2) we do not make any assumption on the true distribution under the alternative hypothesis, and 3) we allow for an asymmetric treatment of false positives (not allowed) and false negatives (limited to $\alpha$ percent). The prior works on hypothesis testing 
with Markovian observation often consider a constrained class of alternative hypotheses, such as fully known (simple) hypothesis~\cite{phatarford1965markov, wald1945SPRT}, or  
a subset of alternatives~\cite{fauss2020minimax,hoeffding1965asymptotically, li2014generalized, zhang2018generalmarkov} with characteristics that can be used to obtain bounds on type-1/2 error in sequential or fixed-length setting. An important class that has received attention recently is the problem of \textit{identity testing}, associated with a contrast function, $d$ and a tolerance $\epsilon$ where the null hypothesis $\boldsymbol{P_X}=\boldsymbol{P}$ is 
tested against \textit{only} the ``$\epsilon$-separated alternatives'' $\{\boldsymbol{P}_X: (\boldsymbol{P},\boldsymbol{P_X})>\epsilon\}$. We also note that many existing works put stringent restrictions on the structure of either the reference Markov chain \cite{fawzi2022sequential}, the true data-generating chain, or both.  

In contrast and motivated by real-life applications in anomaly detection, our work deviates from prior formulation and removes any assumption, including $\epsilon$-distance to the null, on the true distribution under the alternative hypothesis. In fact, the only assumption made in this work is the existence of a stationary distribution for the data-generating chain. Furthermore, considering only the bounded time required to reject the null hypothesis and relaxing the requirement on the tester to accept the null, frees our algorithm and analysis to account for a strongly asymmetric treatment of type-1 versus type-2 errors. More specifically, this allows us to take advantage of a sequential test with a random stopping time which can adapt to the statistical properties of the alternative even when initially nothing was known about it. What is most notable is that our extremely general treatment of alternative composite hypotheses can be essentially done without a loss. 

Our proposed algorithm is motivated and structured similarly to a Markov variant of Wald's original SPRT algorithm~\cite{phatarford1965markov, wald1945SPRT} with the difference that we utilize and deploy a data-driven estimate of the true transition distribution (which is possible with help from Andrew Barron's earlier work on universal estimation of Markov chains \cite{takeuchi2012properties}). Furthermore, we show that by carefully selecting the deployed data-driven estimator, our proposed algorithm is able to asymptotically match that of SPRT in the simple hypothesis testing case with \emph{full knowledge of alternative}. 

While our work utilizes elements of prior work both algorithmically and analytically, due to the inherent differences in formulation, it is challenging to make meaningful comparisons with algorithms proposed in prior work. Nevertheless, we provide detailed simulations results and comparisons with the fixed-length Markov identity test of~\cite{wolfer2020ergodic} in Section~\ref{subsec:adaptivity},
since~\cite{wolfer2020ergodic} allows for the most general class of alternatives in the literature. 



Further afield from our work is the broader topic of property testing, which is thoroughly covered by the survey~\cite{canonne2020survey}.  An important subset of this field is testing if data is generated from an arbitrary distribution with the Markov property~\cite{besag2013exact}, \cite{chen2012testing}.  While this can be used to reject the null when the true distribution is not Markovian, it does not address testing among multiple Markov chains where the alternative is also Markovian.  


%% file: 30_theory.tex
\subsection{General Approach}
We will develop algorithms and analyze their performances for both the i.i.d. and Markov settings.  In both cases our strategy will be to first consider an intuitive modification of the canonical sequential probability ratio test (SPRT)~\cite{wald1945SPRT} for the case of a simple, known alternative hypothesis. In other words, suppose we know that either $X_t \sim p$ or $X_t \sim q$ for some known $p$ and $q$.  The SPRT is  constructed iteratively as 
\begin{align}
\label{eq:statistic}
 L^*_0 & \coloneqq 1,  \\
 \notag L^*_t & \coloneqq L^*_{t-1} \cdot \frac{q(X_t|X_{1:t-1})}{p(X_t|X_{1:t-1})}, \quad \text{for} \quad t \geq 1.
\end{align}
If $p_X =q$ then, in expectation, this statistic will grow over time, so we end the test and declare $H_1$ if ever $L^*_t \geq \nicefrac{1}{\alpha}$ for some fixed parameter $\alpha$, chosen according to the allowed type-1 error rate. 

Following this structure, we will construct a composite test, comparing $p_X{=}p$ to $p_X{\neq} p$, without a known alternative, by combining the following three building blocks.
\begin{description}
    \item[Data-driven Estimator:] In the absence of a known alternative distribution for the observed data, we will need to deploy a causal sequence of empirical estimators, $\{\hat{q}(\cdot|X_{1:t-1}): t \geq 1\}$, of the data-generating distribution. We note that our performance will depend on the choice of empirical estimator $\hat{q}$. Furthermore, there are many existing empirical estimators in the literature. In fact, our analysis identifies an appropriate class of estimators, including one proposed by Takeuchi and Barron \cite{takeuchi2012properties}, discussed in Section~\ref{subsec:estimators}.  
    \item[Empirical Likelihood Ratio:] Given \textit{any} chosen method of generating the estimated distributions, we replace $q$ with $\hat{q}$ in Equation~\ref{eq:statistic} to create an empirical likelihood ratio process:
    \begin{align*}
    {L}_0 & \coloneqq 1,  \\
    {L}_t & \coloneqq {L}_{t-1} \cdot \frac{\hat{q}(X_t|X_{1:t-1})}{p(X_t|X_{1:t-1})}, \quad \text{for} \quad t \geq 1.
\end{align*}
    \item[Threshold-based Stopping Time:] We consider a simple threshold-based stopping time for rejection:
    \[
    \hat{\tau}:= \min\{t: {L}_t \geq 1/\alpha \}.
    \]
\end{description}

Pseudocode for the resulting test applied to Markovian data is given in Algorithm~\ref{alg:markov}, the test for i.i.d. data is the same with the natural substitutions.  

Our alogorithm and test work with any method of estimating the unknown data distribution so long as the resulting $\hat{q}$ is a valid probability distribution: this is characterized by the user-supplied \texttt{GetEstimator} function in Algorithm~\ref{alg:markov}.  This flexibility is a significant advantage of our approach, as there are numerous methods to derive a data-driven estimate of the true distribution and different approaches may be more appropriate for different problem settings. Furthermore, this in general allows for the estimator to account for any side information available about the alternative without needing to modify the test. In the following analysis, we show our
algorithm's performance is contingent on the quality of the estimator and, despite the diversity of ways to characterize an estimator, we identify the precise manner we can quantify this. More specifically, we show that the point-wise cumulative worst-case regret of the log-loss of the estimator is the appropriate metric to assess its performance when used in our setting.  

\begin{definition} 
Let $\hat{q}$ be an empirical (and causal) estimator of the distribution of $X_1, X_2, \ldots$. The cumulative point-wise worst-case regret of the estimator $\{\hat{q}_t(\cdot) = \hat{q}(\cdot|X_{1:t-1}): t \geq 1\}$ is defined as
\begin{small}
\begin{align}
    \label{eq:regret}
    r_t\lp (\hat{q}_i)_{i=1}^t \rp \coloneqq \max_{X_1,\ldots X_t\sim q} \sup_{q \in \Delta_{m-1}} \sum_{i=1}^t \lp \log{\frac{1}{\hat{q}_i(X_i)}} - \log{\frac{1}{q(X_i)}} \rp.
\end{align}
\end{small}
\end{definition}

Note that an estimator with small regret performs well over worst-case choices of the true data distribution it is estimating as well as the sample trajectory drawn from that distribution. 

\subsection{I.I.D.\ Case}
In this setting, we assume that the sequence of samples, $X_1, X_2, \ldots$, are drawn i.i.d. from an unknown distribution $p_X \in \Delta_{m-1}$ and are testing against a known reference distribution, $p \in \Delta_{m-1}$. 


To construct our test for $H_0: p_X = p \text{ vs } H_1: p_X \neq p$ 
we then use an estimate of the true distribution, $\hat{q}_t\in \Delta_{m-1}$. At each time-step $t$ we construct this estimate from the samples observed through the previous time-step: $X_1,\ldots, X_{t-1}$. 

Pseudocode for the Markov version of this test is shown in Algorithm~\ref{alg:markov}, the i.i.d. algorithm is analogous, with the natural substitutions.  This algorithm can be run with any means of constructing the estimates, $\hat{q}_t$, but the performance of the test will depend on the quality of the estimator.  In particular, to obtain our theoretical bounds, we will need estimators with regret $r_t = \mc{O}(\log{t})$.  We discuss in Section~\ref{subsec:estimators} the existence of such estimators and how to choose suitable estimators for different problem settings.  

To analyze the performance of this algorithm we define its stopping time, $\tau \coloneqq \inf\lcb t\in\mathbb{N}  : L_t \geq \nicefrac{1}{\alpha} \rcb$. We then show that it satisfies our desired conditions; under $H_0$ it may end and (incorrectly) reject the null with probability at most $\alpha$ and under $H_1$ it must end and reject the null with probability $1$. 

\begin{theorem}
    \label{thm:iid}
    Given a sequence of i.i.d. observations, this test, using \textbf{any} estimator $\{\hat{q}_t: t \geq 1\}$, has type-1 error at most $\alpha$:
    \begin{align*}
        \P_{H_0}\lp \tau < \infty \rp \leq \alpha.
    \end{align*}

    Furthermore, if the estimator $\{\hat{q}_t: t \geq 1\}$ has a regret guarantee $r_t = \mc{O}\lp \log{t} \rp$, then we also have 
    \begin{align*}
        \E\lb \tau \rb = \mc{O}\lp \frac{\log\lp {\frac{1}{\alpha D(q \parallel p)}}\rp}{D(q \parallel p)} \rp.
    \end{align*}
\end{theorem}

\begin{sproof}
    The details of the proof follow from the proof of the Markov case~\ref{proof:markov}.  At a high level: the proof of the type-1 error bound follows from an application of Ville's Inequality~\ref{subsec:ville}.  Our test statistic forms a non-negative martingale under the null, and Ville provides a bound on the probability of such a martingale ever exceeding a threshold.  

    For the bound on expected stopping time under $H_1$, we exchange our estimated statistic for an oracle statistic based on the true $\boldsymbol{P_X}$, at a price that depends on the regret bound of the estimator.  We then bound the stopping time of the oracle statistic via Wald's identity applied to $\log{L_t}$.  
\end{sproof}
  
\begin{remark}
Note that the type-1 error guarantee is solely due to the non-negative martingale structure of our test statistic. It does not depend on any other detail of the estimator, only that it gives a valid probability distribution.  
\end{remark}

\subsection{Markov Case}
The Markov case proceeds similarly to the i.i.d. case.  Here, our null hypothesis is given by a transition matrix, $\boldsymbol{P}\in\mc{M}^m_1$, and our data consists of a trajectory, $X_1,X_2,\ldots$ drawn from a true, unknown Markov chain given by $\boldsymbol{P_X}\in\mc{M}^m_1$.  We then wish to test 
\begin{align*}
H_0: \boldsymbol{P_X} = \boldsymbol{P}\quad \text{ vs. } \quad H_1: \boldsymbol{P_X} \neq \boldsymbol{P}.
\end{align*}
If we had a known alternative, $\boldsymbol{Q}\in\mc{M}_1^m$, we could construct the SPRT in Equation~\ref{eq:statistic}, with the likelihoods in terms of transition probabilities: $\boldsymbol{Q}(X_t|X_{t-1})$.

In lieu of such an alternative, we again learn an estimator, $\boldsymbol{\hat{Q}_t}\in\mc{M}^m_1$, as we go and use it to construct our test statistic as in Line~\ref{line:statistic} of Algorithm~\ref{alg:markov}.  


\begin{algorithm}[ht]
    \begin{algorithmic}[1]
    \SetAlgoLined
    \STATE \textbf{input:} Null hypothesis $\boldsymbol{P}$, error $\alpha$, initial state $X_0$
    \BlankLine
    \STATE \textbf{initialize:} $L_0 \gets 1$, $t \gets 1$
    \BlankLine
    \WHILE{$L_{t-1} < \nicefrac{1}{\alpha}$}
    \STATE $\boldsymbol{\hat{Q}_t} \gets \texttt{GetEstimator}\big(\lcb X_i \rcb_{i=1}^{t-1}\big)$
    \BlankLine
    \STATE Sample $X_{t} \sim \boldsymbol{P_X}(\cdot|X_{t-1})$
    \BlankLine
    \STATE $L_t \gets L_{t-1} \cdot \frac{\boldsymbol{\hat{Q}_t}(X_t|X_{t-1})}{\boldsymbol{P}(X_t|X_{t-1})}$ \label{line:statistic}
    
    \STATE $t \gets t+1$
    \ENDWHILE
    \RETURN $H_1$
    \end{algorithmic}
    \caption{Test for Markov data}
    \label{alg:markov}
\end{algorithm}

\begin{theorem}
    \label{thm:markov}
    Our Markov test, detailed in Algorithm~\ref{alg:markov}, instantiated with \textbf{any} estimator, has type-1 error at most $\alpha$:
    \begin{align*}
        \P_{H_0}\lp \tau < \infty \rp \leq \alpha.
    \end{align*}
    Furthermore, if the estimator $\{\boldsymbol{\hat{Q}_t}: t \geq 1\}$ has regret guarantee $r_t = \mc{O}\lp \log{t} \rp$, and the true alternative $\boldsymbol{Q}$ is ergodic and satisfies the condition that $\max_{i,j} \boldsymbol{Q}(j|i)/\boldsymbol{P(j|i)} \leq c'$ for some finite constant $c'$,  then
    \begin{align*}
        \E\lb \tau \rb = \mc{O}\lp \frac{\log\lp{\frac{1}{\alpha D_M\lp \boldsymbol{Q} \parallel \boldsymbol{P}\rp }}\rp}{D_M \lp \boldsymbol{Q} \parallel \boldsymbol{P} \rp} \rp.
    \end{align*}
\end{theorem}

\begin{proof}
    \label{proof:markov}
    First we show that $\P_{H_0}(\tau \leq \infty) \leq \alpha$.  

    Let $\{\mc{F}_t: t \geq 0\}$, with $\mc{F}_t\coloneqq \sigma\lp X_1,\ldots, X_t \rp$,  be the natural filtration generated by the samples.  Then note that the test statistic process, $\{L_t: t \geq 0\}$, defines a non-negative martingale since
    \begin{align*}
        \E[L_t|\mc{F}_{t-1}] &= L_{t-1}\E_{\boldsymbol{P}}\lb \boldsymbol{\hat{Q}_t}(X_t|X_{t-1})/\boldsymbol{P}(X_t|X_{t-1})|\mc{F}_{t-1}\rb \\
        &= L_{t-1} \times \sum_{i=j}^m \boldsymbol{\hat{Q}_t}(j|X_{t-1}) = L_{t-1}.
    \end{align*}
    Then, by definition of $\tau$, we have $\lcb \tau < \infty \rcb = \lcb \exists t\geq 0: L_t \geq 1/\alpha \rcb $. An application of Ville's inequality (Appendix~\ref{subsec:ville}) immediately implies that 
    \begin{align*}
        \P_{H_0}\lp \exists t\geq 0:L_t\geq \nicefrac{1}{\alpha}\rp \leq \frac{\E_{H_0}[L_0]}{\nicefrac{1}{\alpha}} = \alpha. 
    \end{align*}

    It then remains to bound the expected stopping time under the alternative, where $\boldsymbol{P_X} {=} \boldsymbol{Q} {\neq} \boldsymbol{P}$.  First define the oracle test statistic with full knowledge of the true Markov chain $\boldsymbol{Q}$:

    \begin{align*}
        L^*_t \coloneqq \prod_{i=1}^t \frac{\boldsymbol{Q}(X_i|X_{i-1})}{\boldsymbol{P}(X_i|X_{i-1})}.
    \end{align*}

    Then we can consider the value of the test statistic, $L_t$, at the stopping time and relate it to the value of the oracle statistic, $L_t^*$, by the control the assumed  regret bound gives on the log loss of $\boldsymbol{\hat{Q}}$ relative to $\boldsymbol{Q}$.  
    \begin{align*}
        \tau &= \inf \lcb t\geq 1: \log{L_t} \geq \log{ \nicefrac{1}{\alpha}} \rcb \\
        &\leq \lcb t\geq 1: \log{L_t^*} \geq \log{\nicefrac{1}{\alpha}}+\frac{m(m-1)}{2}\log{\frac{t}{2\pi}} \rcb \eqqcolon \tau^*.
    \end{align*}

    It follows that $\E[\tau]\leq \E[\tau^*]$, so we proceed to bound $\E[\tau^*]$.

    By definition of $\tau^*$ we have that $\log{L_{\tau^*-1}^*}\leq \log{\nicefrac{1}{\alpha}}+r_{\tau^*-1}$ and by assumption $\E[\log{L_\tau^*}-\log{L_{\tau^*-1}}]$ is bounded a.s. by some constant, $C$.  So we have

    \begin{align}
        \label{eq:bound}
        \E[\log{L_{\tau^*}}] &\leq \log{\nicefrac{1}{\alpha}}+\E \lb \log{\tau^*} \rb + C\\
        \notag&\leq \log{\nicefrac{1}{\alpha}}+\log{\E[\tau^*]} +C,
    \end{align}
    accumulating constants independent of $t$ and applying Jensen's inequality for the second line.  

    By a Markov chain version of Wald's identity, due to~\cite{moustakides1999markovwald} and detailed in Appendix~\ref{subsec:wald}, we have that 
    \begin{align}
        \label{eq:wald}\E[\log{L^*_{\tau^*}}]&= \E\lb \sum_{i=1}^{\tau^*}\log{\frac{\boldsymbol{Q}(X_i|X_{i-1})}{\boldsymbol{P}(X_i|X_{i-1})}} \rb  \\
        \notag &= D_M(\boldsymbol{Q}\parallel \boldsymbol{P})\cdot \E[\tau^*] + C'.
    \end{align}

Combining equations~\ref{eq:bound} and~\ref{eq:wald} yields

\begin{align*}
    D_M(\boldsymbol{Q}\parallel \boldsymbol{P})\cdot \E[\tau^*] \leq \log{\nicefrac{1}{\alpha}} + \log{\E[\tau^*]} + \mc{O}(1)
\end{align*}
and simplification gives the desired bound. 
\end{proof}
\begin{remark}
The one-sided nature of our problem setting facilitates the use of the estimated distribution here.  Loosely speaking, under the alternative, where $\boldsymbol{P_X}{=}\boldsymbol{Q}{\neq} \boldsymbol{P}$, we expect $\boldsymbol{\hat{Q}_t}$ to converge to $\boldsymbol{Q}$ as $t$ increases. This then will bias the test statistic upwards as, in expectation, we'll have $\nicefrac{\boldsymbol{\hat{Q}_t}(X_t|X_{t-1})}{\boldsymbol{P}(X_t|X_{t-1})} {>} 1$.  So eventually the statistic will exceed $\nicefrac{1}{\alpha}$ and reject the null.  

On the other hand, under $H_0$ we expect that $\boldsymbol{\hat{Q}_t}$ will converge to $\boldsymbol{P}$ and so the statistic will stagnate.  For a two-sided test this would be a problem, as we would need to find some criteria under which to accept the null.  But in our setting the desired behavior under the null is to continue the test indefinitely, eliminating this challenge.  
\end{remark}

\begin{remark}
    The bounds on our algorithms' expected stopping times depend on the distance between our known null hypotheses and the \textbf{unknown} alternative.  This means that, when executing a test, we cannot calculate the bound a priori, as we don't know $\boldsymbol{Q}$.  But this is because the sequential nature of our algorithm allows it to adapt to the unknown difficulty of the problem.  When $\boldsymbol{Q}$ and $\boldsymbol{P}$ are very distinct, as measured by $D_M(\cdot {\parallel} \cdot )$, the expected stopping time is small as the algorithm is able to quickly distinguish the observed samples from the null.  In contrast, a fixed-length test has to ensure sufficient samples to accommodate worst-case problem instances.     
\end{remark}

\begin{remark}
    The assumption on the likelihood ratio; that is, $\max_{i,j} \boldsymbol{Q}(j|i)/\boldsymbol{P}(j|i) < c'$ for an unknown but finite constant $c'$, was made in the statement of~Theorem~\ref{thm:markov} mainly to simplify the analysis. A similar result can be proved without this bounded likelihood assumption, but with a slightly more involved argument.
\end{remark}

\subsection{Estimators}
\label{subsec:estimators}
The key element of our algorithm is the ability to estimate a distribution from its samples.  This problem has been studied in many different settings for the i.i.d. case, see~\cite{Kamath2015samples} and references therein for further discussion.  

As discussed above, our bound on expected stopping time depends on the use of an estimator with bounded pointwise log-loss regret~\ref{eq:regret}.  

In the i.i.d. case the Krichevsky-Trofimov (KT, ``add-$\nicefrac{1}{2}$'') estimator~\ref{eq:kt} was shown to have regret $\mc{O}\lp \frac{m-1}{2} \log \lp  \frac{t}{2\pi} \rp \rp$~\cite{krichevsky1981kt}.

\begin{align}
    \label{eq:kt}
    \hat{q}_{t+1}(x) = \frac{\sum_{i=1}^{t}\indi{X_i=x}+\nicefrac{1}{2}}{t+\nicefrac{m}{2}}.
\end{align}

This estimator also has a Bayesian interpretation: it results from the choice of the \textit{Jeffreys prior}, a non-informative prior for a model's parameter space constructed from the Fisher information of the model.  For the parameters of a categorical distribution the Jeffreys prior is a Dirichlet-$\nicefrac{1}{2}$ $(D_{(1/2)})$ distribution.  

Optimal estimation is more involved in the Markov setting as it involves not only the process of estimating the conditional distribution of each state, but the reachability of each state: it may take a very long time to get sufficient samples to estimate the conditional distribution of rarely seen states.  And as our results depend on pointwise regret guarantees we cannot rely on the rarity of these states to mitigate their impact.  This dependency can be seen even in those identity testing algorithms that don't depend on estimation, e.g. in the mixing time dependency in the complexity of the algorithm in~\cite{wolfer2020ergodic}.   

This effect is also apparent in the use of Jeffreys mixture for sequential estimation, which is shown to be optimal in~\cite{takeuchi2012properties}.  This approach estimates the true, unknown Markov chain by a weighted mixture of all first order ergodic Markov chains.  In particular it weights each $\boldsymbol{Q}\in\mc{M}^m_1$ with stationary distribution $\rho$ according to Jeffreys prior, defined up to normalization in Equation~\ref{eq:jeffreys}, and performs Bayesian updates to the mixture weight with each acquired sample.  

\begin{align}
\label{eq:jeffreys}
    \pi(\boldsymbol{Q}) \propto \prod_{i=1}^m \rho(i)^\frac{m-1}{2}D_{(\nicefrac{1}{2})}(\boldsymbol{Q}(\cdot|i))
\end{align}

This amounts to putting a $D_{(\nicefrac{1}{2})}$ prior on each state's transition distribution, weighted by the stationary distribution to address the concerns raised above.  But this weighting also has the effect of entangling the estimation of the distribution for each state, resulting in Bayesian update formula without the simple form of the KT estimator. Prediction with Jeffreys mixture in this setting is studied extensively in~\cite{takeuchi2012properties}, where the authors show it attains regret $r_t(\boldsymbol{\hat{Q}_t^J}) = \mc{O}\lp \frac{(m-1)m}{2}\log{\frac{t}{2\pi}}\rp$ and derive efficient approximations for the Bayesian updates.  

While this suffices for both the implementation of our algorithm and the proof of Theorem~\ref{thm:markov}, their main result is that Jeffreys mixture is not asymptotically minimax optimal: it places too little mass on the edges of $\mc{M}_1^m$--i.e. Markov chains with some transition probabilities near 0.  This deficiency can be remedied by instead using a density given by the weighted sum of Jeffreys mixture and a density which places more mass on sparse elements of $\mc{M}_1^m$.  This modified estimator is not compatible with a sequential test, as it requires foreknowledge of the total number of samples, but in the anomaly detection setting we are concerned with rare events, and so chains with very small transition probabilities, so we do investigate the impacts of these results in Section~\ref{subsec:estimators}.  

Finally, while these estimators are suitable for our theoretical analysis, the design of our algorithm makes it easily compatible with any means of estimating a distribution.  This leads to a variety of interesting directions for practical improvements and implementations in settings lacking a simple Markovian structure, which we discuss in Section~\ref{subsec:future}. 

%% file: 40_simulations.tex
\subsection{Sequential Adaptivity}
\label{subsec:adaptivity}
To illustrate the value of sequential testing in this setting, we ran simulations comparing our sequential test to the fixed-length test of Wolfer and Kontorovich~\cite{wolfer2020ergodic}.  This test takes a trajectory sampled from an unknown Markov chain, $\boldsymbol{P_X}$, and an ergodic reference chain, $\boldsymbol{P}$, and decides whether $H_0: \boldsymbol{P}=\boldsymbol{P_X}$ or 
\begin{align*}
H_1: \norm{\boldsymbol{P}-\boldsymbol{P_X}}_M\coloneqq  2\max_{i\in[m]}\norm{(\boldsymbol{P}-\boldsymbol{P_X})(\cdot|i)}_{\text{TV}}>\epsilon
\end{align*}

with high probability.  Note that the behavior of the test is undefined when $\norm{\boldsymbol{P_X}-\boldsymbol{P}}_M\leq \epsilon$, but $\boldsymbol{P_X}\neq \boldsymbol{P}$. 

The test first checks if the frequency of states in the trajectory is sufficiently similar to $\boldsymbol{P}$'s stationary distribution, if it then each of the empirical transition distributions are compared against the rows of $\boldsymbol{P}$ using the i.i.d. identity tester of~\cite{valiant2017automatic}.

This algorithm is designed for the fixed length setting, while ours is sequential. So this is not a comparison of the efficacy or efficiency of the algorithms, but meant only to demonstrate the ability of our sequential algorithm to automatically adapt to the relative difficulty of the problem.    

\begin{figure}[t]
\centerline{\includegraphics[width=\columnwidth]{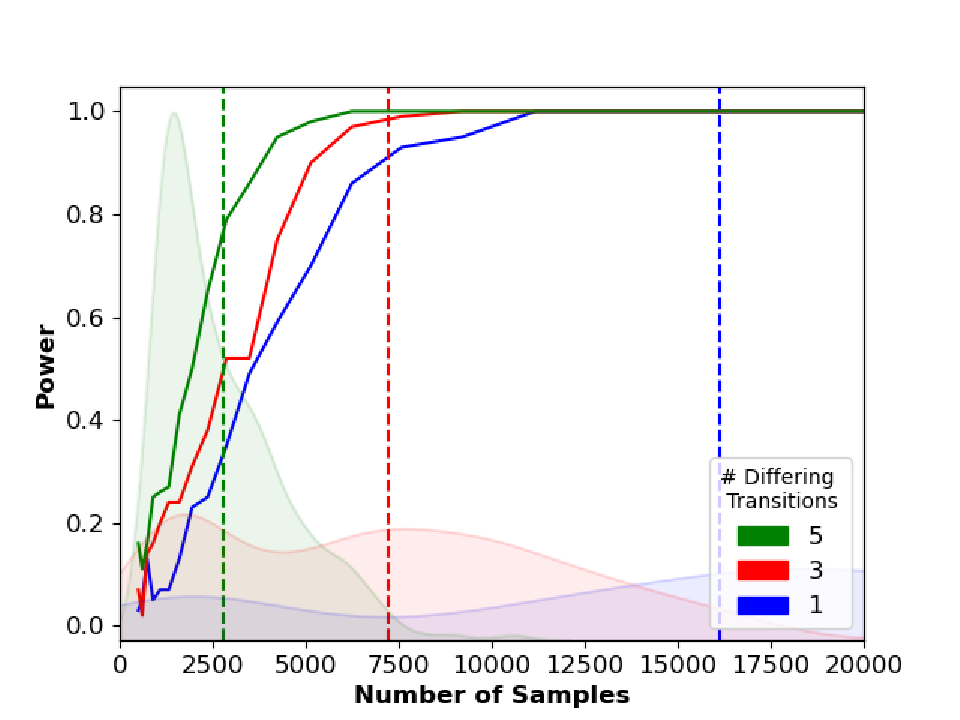}}
\caption{Power of the fixed length test of Wolfer and Kontorovich~\cite{wolfer2020ergodic}, shown by the solid curves, relative to the expected stopping time of our sequential test, marked by the dashed line, with the distribution of stopping times in the background. }
\label{fig:stopping_times}
\end{figure}

To run these simulations we first generated a random reference $m{=}5$-state chain, $\boldsymbol{P}$.  We then ran the fixed-length algorithm with $\boldsymbol{P_X} = \boldsymbol{P}$ for 500 trials to find the average type-1 error of the test.  Using this error level to set $\alpha$ in Algorithm~\ref{alg:markov}, we then ran our sequential algorithm on a randomly generated alternative with $\norm{\boldsymbol{P_X}-\boldsymbol{P}}_M>\epsilon$.  The stopping time for several problem instances is shown in Figure~\ref{fig:stopping_times}, with the average over 500 trials given by the dashed lines.  The solid curves show the power of the fixed length test: using the same alternatives we ran the fixed length test for a range of sample sizes and recorded the empirical false accept rate over 500 trials for each sample size. Note that this comparison is imperfect as the type 1 error of the fixed length test could also vary with sample size, but in practice, for the range of samples displayed in Figure~\ref{fig:stopping_times}, the type 1 error remained consistent, around $5\%$ with the settings used, while the additional samples improved the power of the test.  

To illustrate the adaptivity of our sequential algorithm we created problem instances of varying difficulty.  To do this we took the randomly generated $\boldsymbol{P}$ and created the hidden data generating chain, $\boldsymbol{P_X}$, by randomly perturbing varying numbers of rows of $\boldsymbol{P}$.  We ensured that at least one perturbation gave $\norm{\boldsymbol{P}(\cdot|i)-\boldsymbol{P_X}(\cdot|i)}_{\text{TV}}>\nicefrac{\epsilon}{2}$, so that it forms a valid alternative in the fixed length test setting.  Then, as more rows are perturbed, the test becomes easier since the observed samples are being drawn from a chain more distinguishable from the null hypothesis.  

The results show that, as expected, both algorithms take significantly more samples to perform well on the harder problem instances.  The important distinction here, though, is that the fixed length algorithm has to be given a trajectory of some set length without knowing the difficulty of the problem--meaning that to guarantee a power 1 test in this setting, the fixed length test always requires well over $10,000$ samples despite the fact that it can achieve good performance with a fraction of the samples on easier problem instances.  The strength of sequential algorithms in general, and ours here, is that they can adapt to easier problem instances without prior knowledge, automatically terminating the test upon achieving sufficient confidence.  

\subsection{Estimator Evaluation}

In this subsection, we provide a simple numerical example to highlight how the choice of specific data-driven estimators impacts the non-asymptotic performance of our proposed algorithm. 

We note that Jeffreys mixture does not result in an optimal regret as shown by ~\cite{takeuchi2012properties} where the proposed modification to Jeffreys mixture was derived and proved to be optimal in the worst-case regret sense. More precisely,~\cite{takeuchi2012properties} shows that 
Jeffreys mixture itself puts too little mass on those Markov chains near the edges of $\mc{M}^m_1$, the chains which have some transition probabilities near 0.  This deficiency is exacerbated by some of the common add-a-constant estimators.  The add-$\nicefrac{1}{2}$ and add-$1$ estimators are most common and, in the Markov case, are equivalent to putting an independent $D_{(\nicefrac{1}{2})}$ and $D_{(1)}$ prior, respectively, on each row of the transition matrices, and larger parameters of a Dirichlet bias the estimator away from sparse distributions. 

Our experimental results, however, suggest that these computationally hard to realize modifications to Jeffreys mixture are not necessary to achieve good performance in our sequential setting. In this section, we aim to evaluate the difference between some of these estimators. In particular, we will provide a simple example to quantify the performance difference between 1) the optimal modified Jeffreys mixture, 2) Jeffreys mixture, and 3) the add-$\nicefrac{1}{2}$ estimator. We ran our algorithm, instantiated with each of these estimators, on the toy problem~\ref{eq:toy_problem} with the transition matrices



\begin{small}
\begin{align}
    \label{eq:toy_problem}
        \boldsymbol{P} = \begin{pmatrix}
        \epsilon &1-\epsilon \\
        0.7      &0.3
    \end{pmatrix} \quad \text{vs} \quad
    \boldsymbol{P_X} = 
    \begin{pmatrix}
        \epsilon &1-\epsilon \\
        0.9      &0.1
    \end{pmatrix}. 
\end{align}
\end{small}

This problem is designed to test the quality of the estimators near an edge of $\mc{M}^2_1$, where as $\epsilon\to 0$ we expect the (non-optimal) estimators, besides the modified Jeffreys mixture, to exhibit some bias away from the true sparse, unknown, $\boldsymbol{P_X}.$   

\begin{figure}[h]
\centerline{\includegraphics[width=0.8\columnwidth]{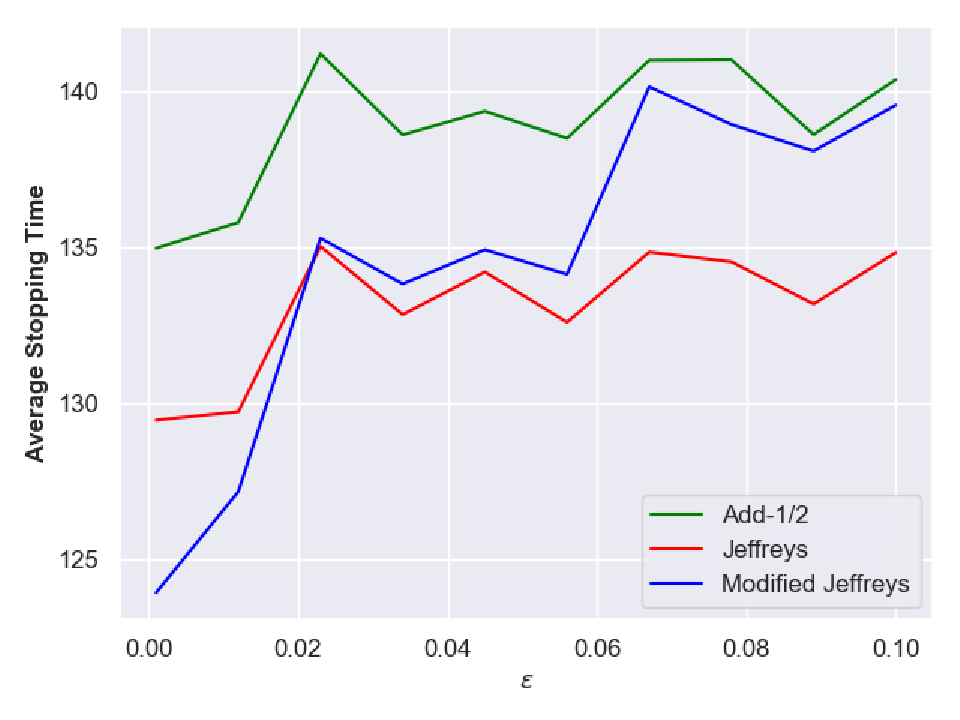}}
\caption{Average stopping time of our algorithm instantiated with several different estimators for Problem~\ref{eq:toy_problem} as $\epsilon$ varies.}
\label{fig:estimators}
\end{figure}

Figure~\ref{fig:estimators} shows the resulting stopping times for our algorithm with each estimator as a function of $\epsilon$.  We bring attention to the following observations: 
\begin{itemize}
    \item Jeffreys mixture is uniformly better than the add-$\nicefrac{1}{2}$ estimator; however, this gap is not significant in the short-term. 
    \item As $\epsilon$ gets very small, the modified Jeffreys mixture noticeably outperform the non-optimal estimators; however, the improvements are both computationally expensive and are not robust in terms of choice of estimator parameter.   
\end{itemize} 

These observations suggest that the need for the more elaborate estimator is tied to our interest not to restrict the alternative hypothesis. An important question for future work is the interplay between side information/assumption on alternative hypothesis and the complexity of the estimator. This is of particular interest in the sequential setting where an adaptive choice of estimator might be possible. 

%% file: 50_conclusion.tex
\label{subsec:future}
As discussed in~\ref{subsec:related}, a wide variety of related problem settings have been studied in many different fields.  Drawing connections between these different settings would be valuable for understanding the trade-offs implicit in, for instance, our choice to allow the test to run indefinitely under the null hypothesis instead of allowing an uncertainty ball around the null hypothesis, as in the identity testing literature.  Rigorous comparison between these settings, or a sequential algorithm developed natively for the identity testing formulation, would also allow a clear characterization of the performance improvements sequential algorithms offer over their fixed-length counterparts.    

An interesting extension of our work lays in the somewhat stringent assumption that the data is generated by a first-order Markov chain.  Estimators for higher, but known, order Markov chains are also discussed in~\cite{takeuchi2012properties} and could easily be fit into our algorithm.  Relaxing our assumptions further, Bayesian context trees~\cite{kontoyiannis2022bayesian} offer estimators that could be applied to data with variable-length memory.  And finally, the use of recurrent neural networks to learn a context-dependent estimator, as in~\cite{salinas2020deepar} could allow implementation of our algorithm in settings with unspecified, arbitrary dependency structures.  

We also presently rely on explicit knowledge of the null hypothesis dynamics.  In practice we may also need to estimate this, perhaps from a period of known optimal behavior, before beginning our test.  Here we may be able to leverage results from the rich field of change point detection, which covers a variety of methods intended to alert upon a change in the model generating a stream of data and is surveyed in~\cite{aminikhanghahi2017changesurvey} and studied specifically in a Markov context in~\cite{lungu2022treechange} and~\cite{fuh2004asymptoticchange} among others.  

%% file: 90_appendix.tex
\section{}
\subsection{Ville's Inequality}
\label{subsec:ville}
\begin{theorem}{Ville (1939)}
    For a non-negative supermartingale, $L_1,L_2,\ldots$ and any $\beta>1$, define the stopping time
    \begin{align*}
        \tau \coloneqq \inf \lcb t\geq 1: L_t \geq \beta \rcb.
    \end{align*}
    Then 
    \begin{align*}
        \P(\tau < \infty) = \P \big( \lcb \exists t: L_t \geq \beta \rcb \big) \leq \frac{\E[L_0]}{\beta}. 
    \end{align*}
\end{theorem}

\subsection{Wald's Identity for Markov Chains}
\label{subsec:wald}
The key difference in the proofs of the i.i.d. and Markov case is in the extension of Wald's well-known identity to the Markov case.  A satisfactory extension for our purposes is given by~\cite{moustakides1999markovwald}--see sections 4.1 and 5 therein for a detailed discussion of the points we summarize here.

Wald's identity gives that, if $X_0, X_1, \ldots$ is a sequence of i.i.d. realizations of a finite mean random variable and $\tau$ is a finite expectation stopping time, and $\theta$ is an appropriately bounded function, then $\E[\sum_{i=0}^\tau \theta(X_i)] = \E[\tau]\E[\theta(X_0)]$.  The Markov version reproduces this result, plus a drift term that depends on how closely the chain adheres to its stationary distribution.  This drift term is defined by the solution to a constrained Poisson integral equation.  In the case of a discrete state space this simplifies considerably to the solution of the linear equations 

\begin{align*}
    (\boldsymbol{P}-\mathbb{I})\omega &= -(\boldsymbol{P}-J\pi^T)\theta, \\
    \pi^T\omega &= 0,
\end{align*}

where $J=(1...1)^T$.  

The existence of a solution to these equations requires some assumptions, detailed in Theorem 1.1 of~\cite{moustakides1999markovwald}.  In the finite state space we can give sufficient conditions on the eigenvalues of $\boldsymbol{P}$: if the unit eigenvalue is simple and the magnitudes of all other eigenvalues are strictly less than one then there is a uniformly bounded solution, $\omega$.  This allows us to disregard the drift term, detailed in Lemma~\ref{lemma:wald_markov}, in the asymptotic regime.  

Given the solution to this system, the Markov Wald's identity is then   
\begin{lemma}
    \label{lemma:wald_markov}
    \begin{align*}
        \E\lb \sum_{t=0}^\tau \theta(X_i) \rb = \E[\tau]\lim_{n\to \infty} \E[\theta(X_n)] + \E\lb \omega(X_0)-\omega(X_\tau) \rb.
    \end{align*}
\end{lemma}

The function of interest for our purposes is $\theta(X_i, X_{i-1}) \coloneqq \log{\frac{\boldsymbol{Q}(X_i|X_{i-1})}{\boldsymbol{P}(X_i|X_{i-1})}}$, which is a function of two states of the Markov chain.  But, since our stopping rule is adapted to the filtration generated by the sample trajectory, we can instead define

\begin{align*}
    \theta(X_{i-1}) = \E_{X_{i}} \lb \log{\frac{\boldsymbol{Q}(X_i|X_{i-1})}{\boldsymbol{P}(X_i|X_{i-1})}} \bigg| X_{i-1} \rb 
\end{align*}

and then we have that 

\begin{align*} 
    \E \lb  \sum_{i=1}^\tau \log{\frac{\boldsymbol{Q}(X_i|X_{i-1})}{\boldsymbol{P}(X_i|X_{i-1})}} \rb {=} \E\lb \sum_{i=1}^\tau \theta(X_i)\rb.
\end{align*}

To which we can apply Lemma~\ref{lemma:wald_markov}.  It remains to evaluate $\lim_{n\to \infty} \E[\theta(X_n)]$.  The above assumption on the eigenvalues of the data generating chain--here $\boldsymbol{Q}$--also gives aperiodicity and so convergence to a stationary distribution, $\rho$.  Thus we have
\begin{align*}
    \lim_{n\to\infty}\mathbb{E}[\theta(X_n)] &= \lim_{n\to\infty} \sum_{i\in[m]}\mathbb{P}(X_n{=}i)\sum_{j\in[m]}\boldsymbol{Q}(j|i)\log{\frac{\boldsymbol{Q}(j|i)}{\boldsymbol{P}(j|i)}} \\
    &=\sum_{i\in[m]}\rho_i\sum_{j\in[m]} \boldsymbol{Q}(j|i)\log{\frac{\boldsymbol{Q}(j|i)}{\boldsymbol{P}(j|i)}} \\
    &= D_M(\boldsymbol{Q} \parallel \boldsymbol{P} )
\end{align*}
as used in the proof of Theorem~\ref{thm:markov}.